# Lower bounds for the greatest possible number of colors in interval edge colorings of bipartite cylinders and bipartite tori

Petros A. Petrosyan

Institute for Informatics and Automation Problems of NAS of RA,
Department of Informatics and Applied Mathematics, YSU,
Yerevan, Armenia
e-mail: pet_petros@ipia.sci.am,
pet_petros@yahoo.com

Gagik H. Karapetyan

Department of Informatics and Applied Mathematics, YSU,
Yerevan, Armenia
e-mail: kar_gagik@yahoo.com

**ABSTRACT**

An interval edge $t$-coloring of a graph $G$ is a proper edge coloring of $G$ with colors $1, 2, \ldots, t$ such that at least one edge of $G$ is colored by color $i$, $i = 1, 2, \ldots, t$, and the edges incident with each vertex $v \in V(G)$ are colored by $d_G(v)$ consecutive colors, where $d_G(v)$ is the degree of the vertex $v$ in $G$. In this paper interval edge colorings of bipartite cylinders and bipartite tori are investigated.

## 1. INTRODUCTION

All graphs considered in this paper are finite, undirected and have no loops or multiple edges. Let $V(G)$ and $E(G)$ denote the sets of vertices and edges of a graph $G$, respectively. The degree of a vertex $v \in V(G)$ is denoted by $d_G(v)$, the maximum degree of a vertex of $G$ - by $\Delta(G)$, the chromatic index of $G$ - by $\chi'(G)$, and the diameter of $G$ - by $d(G)$. Given two graphs $G_1 = (V_1, E_1)$ and $G_2 = (V_2, E_2)$, the Cartesian product $G_1 \times G_2$ is a graph $G = (V, E)$ with vertex set $V = V_1 \times V_2$ and the edge set $E = \{((u_1, u_2), (v_1, v_2)) \mid \text{either } u_1 = v_1 \text{ and } (u_2, v_2) \in E_2 \text{ or } u_2 = v_2 \text{ and } (u_1, v_1) \in E_1\}$. The bipartite cylinder $C(m, 2n)$ is the Cartesian product $P_m \times C_{2n}$ ($m \in N$, $n \geq 2$) and the bipartite torus $T(2m, 2n)$ is the Cartesian product $C_{2m} \times C_{2n}$ ($m \geq 2$, $n \geq 2$). If $\alpha$ is a proper edge coloring of the graph $G$ then $\alpha(e)$ denotes the color of an edge $e \in E(G)$ in the coloring $\alpha$. For a proper edge coloring $\alpha$ of a graph $G$ and for any $v \in V(G)$ we denote by $S(v, \alpha)$ the set of colors of edges incident with $v$.

An interval [1] edge $t$-coloring of a graph $G$ is a proper edge coloring of $G$ with colors $1, 2, \ldots, t$ such that at least one edge of $G$ is colored by color $i$, $i = 1, 2, \ldots, t$, and the edges incident with each vertex $v \in V(G)$ are colored by $d_G(v)$ consecutive colors.

For $t \geq 1$ let $\mathfrak{N}_t$ denote the set of graphs which have an interval edge $t$-coloring, and assume: $\mathfrak{N} \equiv \bigcup_{t \geq 1} \mathfrak{N}_t$. For a graph $G \in \mathfrak{N}$ the least and the greatest values of $t$, for which $G \in \mathfrak{N}_t$, are denoted by $w(G)$ and $W(G)$, respectively.

The problem of deciding whether or not a bipartite graph belongs to $\mathfrak{N}$ was shown in [2] to be $NP$-complete [3,4].

It was proved in [5] that if $G = C(m, 2n)$ or $G = T(2m, 2n)$ then $G \in \mathfrak{N}$ and $w(G) = \Delta(G)$.

**Theorem 1** [6]. If $G$ is a bipartite graph and $G \in \mathfrak{N}$ then $W(G) \leq d(G)(\Delta(G) - 1) + 1$.

**Theorem 2** [7]. Let $G$ be a regular graph.
1. $G \in \mathfrak{N}$ iff $\chi'(G) = \Delta(G)$.
2. If $G \in \mathfrak{N}$ and $\Delta(G) \leq t \leq W(G)$ then $G \in \mathfrak{N}_t$.

In this paper interval edge colorings of bipartite cylinders and bipartite tori are investigated. The terms and concepts that we do not define can be found in [8-10].

## 2. LOWER BOUNDS FOR $W(C(m, 2n))$ AND $W(T(2m, 2n))$.

**Theorem 3.** If $G = C(m, 2n)$ then $W(G) \geq 3m + n - 2$.
**Proof.** Let
$$V(G) = \{x_j^{(i)} \mid 1 \leq i \leq m, 1 \leq j \leq 2n\},$$
$$E(G) = \left(\bigcup_{i=1}^{m} E^i(G)\right) \cup \left(\bigcup_{j=1}^{2n} E_j(G)\right),$$
where
$$E^i(G) = \{(x_j^{(i)}, x_{j+1}^{(i)}) \mid 1 \leq j \leq 2n-1\} \cup \{(x_1^{(i)}, x_{2n}^{(i)})\},$$
$$E_j(G) = \{(x_j^{(i)}, x_j^{(i+1)}) \mid 1 \leq i \leq m-1\}.$$

Define an edge coloring $\alpha$ of the graph $G$ in the following way:
1. for $i = 1, 2, \ldots, m$, $j = 1, 2, \ldots, n+1$
$$\alpha\left((x_j^{(i)}, x_{j+1}^{(i)})\right) = 3i + j - 3;$$
2. for $i = 1, 2, \ldots, m$, $j = n+2, \ldots, 2n-1$
$$\alpha\left((x_j^{(i)}, x_{j+1}^{(i)})\right) = 3i - j + 2n - 1;$$
3. for $i = 1, 2, \ldots, m$
$$\alpha\left((x_1^{(i)}, x_{2n}^{(i)})\right) = 3i - 1;$$

4. for $i = 1, 2, \ldots, m-1$, $j = 2, 3, \ldots, n+1$
$$\alpha\left(\left(x_j^{(i)}, x_j^{(i+1)}\right)\right) = 3i + j - 2;$$
5. for $i = 1, 2, \ldots, m-1$, $j = n+2, \ldots, 2n$
$$\alpha\left(\left(x_j^{(i)}, x_j^{(i+1)}\right)\right) = 3i - j + 2n + 1;$$
6. for $i = 1, 2, \ldots, m-1$
$$\alpha\left(\left(x_1^{(i)}, x_1^{(i+1)}\right)\right) = 3i.$$

Let us show that $\alpha$ is an interval edge $(3m+n-2)$-coloring of the graph $G$.

First of all let us prove that for $i$, $i = 1, 2, \ldots, 3m+n-2$ there is an edge $e_i \in E(G)$ such that $\alpha(e_i) = i$.

For $i = 1, 2, \ldots, m$ we define a set $F_i$ in the following way:
$$F_i = \left\{\alpha\left(\left(x_j^{(i)}, x_{j+1}^{(i)}\right)\right) \mid 1 \leq j \leq n+1\right\}.$$

Clearly,
$$F_i = \{3i-2, 3i-1, \ldots, 3i+n-2\}, \quad |F_i| = n+1, \quad i = 1, 2, \ldots, m.$$

It is not hard to check that
$$\bigcup_{i=1}^{m} F_i = \{1, 2, \ldots, 3m+n-2\},$$

and, therefore for $i$, $i = 1, 2, \ldots, 3m+n-2$ there is an edge $e_i \in E(G)$ such that $\alpha(e_i) = i$.

Now, let us show that the edges that are incident to a vertex $v \in V(G)$ are colored by $d_G(v)$ consecutive colors.

Let $x_j^{(i)} \in V(G)$, where $1 \leq i \leq m, 1 \leq j \leq 2n$.

Case 1. $i = 1$, $j = 1, 2$.
It is not hard to see that
$$S\left(x_j^{(i)}, \alpha\right) = \{3i-2, 3i-1, 3i\}.$$

Case 2. $i = 1$, $j = 3, \ldots, 2n$.
It is not hard to see that
$$S\left(x_k^{(i)}, \alpha\right) = S\left(x_{2n+3-k}^{(i)}, \alpha\right) = \{k-1, k, k+1\},$$
where $k = 3, \ldots, n+1$.

Case 3. $i = m$, $j = 1, 2$.
It is not hard to see that
$$S\left(x_j^{(i)}, \alpha\right) = \{3i-3, 3i-2, 3i-1\}.$$

Case 4. $i = m$, $j = 3, \ldots, 2n$.
It is not hard to see that
$$S\left(x_k^{(i)}, \alpha\right) = S\left(x_{2n+3-k}^{(i)}, \alpha\right) = \{3i+k-5, 3i+k-4, 3i+k-3\},$$
where $k = 3, \ldots, n+1$.

Case 5. $i = 2, 3, \ldots, m-1$, $j = 1, 2$.
It is not hard to see that
$$S\left(x_j^{(i)}, \alpha\right) = \{3i-3, 3i-2, 3i-1, 3i\}.$$

Case 6. $i = 2, 3, \ldots, m-1$, $j = 3, \ldots, 2n$.
It is not hard to see that
$$S\left(x_k^{(i)}, \alpha\right) = S\left(x_{2n+3-k}^{(i)}, \alpha\right) = \{3i-k+2n-2, 3i-k+2n-1,$$
$$3i-k+2n, 3i-k+2n+1\}, \text{ where } k = n+2, \ldots, 2n.$$

Therefore, $\alpha$ is an interval edge $(3m+n-2)$-coloring of the graph $G$.
The proof is complete.

**Remark.** Since $C(m, 2n)$ is a bipartite graph with $2 \leq \Delta(C(m, 2n)) \leq 4$ and $d(C(m, 2n)) = m + n - 1$ then from **theorem 1** we have $W(C(m, 2n)) \leq 3m + 3n - 2$.

**Theorem 4.** If $G = T(2m, 2n)$ then
$$W(G) \geq \max\{3m+n, 3n+m\}$$

**Proof.** Let $m \leq n$ and $V(G) = \left\{x_j^{(i)} \mid 1 \leq i \leq 2m, 1 \leq j \leq 2n\right\}$,
$$E(G) = \left(\bigcup_{i=1}^{2m} E^i(G)\right) \cup \left(\bigcup_{j=1}^{2n} E_j(G)\right), \text{ where}$$
$$E^i(G) = \left\{\left(x_j^{(i)}, x_{j+1}^{(i)}\right) \mid 1 \leq j \leq 2n-1\right\} \cup \left\{\left(x_1^{(i)}, x_{2n}^{(i)}\right)\right\},$$
$$E_j(G) = \left\{\left(x_j^{(i)}, x_j^{(i+1)}\right) \mid 1 \leq i \leq 2m-1\right\} \cup \left\{\left(x_j^{(1)}, x_j^{(2m)}\right)\right\}.$$

Define an edge coloring $\beta$ of the graph $G$ in the following way:
1. for $i = 1, 2, \ldots, m$, $j = 1, 2, \ldots, n+1$
$$\beta\left(\left(x_j^{(i)}, x_{j+1}^{(i)}\right)\right) = \beta\left(\left(x_j^{(2m+1-i)}, x_{j+1}^{(2m+1-i)}\right)\right) = i + 3j - 3;$$
2. for $i = 1, 2, \ldots, m$, $j = n+2, \ldots, 2n-1$
$$\beta\left(\left(x_j^{(i)}, x_{j+1}^{(i)}\right)\right) = \beta\left(\left(x_j^{(2m+1-i)}, x_{j+1}^{(2m+1-i)}\right)\right) = i - 3j + 6n + 3;$$
3. for $i = 1, 2, \ldots, m$
$$\beta\left(\left(x_1^{(i)}, x_{2n}^{(i)}\right)\right) = \beta\left(\left(x_1^{(2m+1-i)}, x_{2n}^{(2m+1-i)}\right)\right) = i + 3;$$
4. for $i = 1, 2, \ldots, m$, $j = 2, 3, \ldots, n+1$
$$\beta\left(\left(x_j^{(i)}, x_j^{(i+1)}\right)\right) = \beta\left(\left(x_j^{(2m-i)}, x_j^{(2m+1-i)}\right)\right) = i + 3j - 4;$$
5. for $i = 1, 2, \ldots, m$, $j = n+2, \ldots, 2n$
$$\beta\left(\left(x_j^{(i)}, x_j^{(i+1)}\right)\right) = \beta\left(\left(x_j^{(2m-i)}, x_j^{(2m+1-i)}\right)\right) = i - 3j + 6n + 5;$$
6. for $i = 1, 2, \ldots, m$
$$\beta\left(\left(x_1^{(i)}, x_1^{(i+1)}\right)\right) = \beta\left(\left(x_1^{(2m-i)}, x_1^{(2m+1-i)}\right)\right) = i + 2;$$
7. for $j = 3, \ldots, n+1$
$$\beta\left(\left(x_j^{(1)}, x_j^{(2m)}\right)\right) = \beta\left(\left(x_{2n+3-j}^{(1)}, x_{2n+3-j}^{(2m)}\right)\right) = 3j - 4;$$
8. $\beta\left(\left(x_1^{(1)}, x_1^{(2m)}\right)\right) = \beta\left(\left(x_2^{(1)}, x_2^{(2m)}\right)\right) = 2.$

Let us show that $\beta$ is an interval edge $(3n+m)$-coloring of the graph $G$.

First of all let us prove that for $i$, $i = 1, 2, \ldots, 3n+m$ there is an edge $e_i \in E(G)$ such that $\beta(e_i) = i$.

It is not hard to check that
$$\bigcup_{i=1}^{m} \bigcup_{j=1}^{n+1} S\left(x_j^{(i)}, \beta\right) = \{1, 2, \ldots, 3n+m\},$$

and, therefore for $i$, $i = 1, 2, \ldots, 3n+m$ there is an edge $e_i \in E(G)$ such that $\beta(e_i) = i$.

Now, let us show that the edges that are incident to a vertex $v \in V(G)$ are colored by four consecutive colors.

Let $x_j^{(i)} \in V(G)$, where $1 \leq i \leq 2m, 1 \leq j \leq 2n$.

Case 1. $i = 1, 2, \ldots, 2m$, $j = 1$.
It is not hard to see that
$$S\left(x_j^{(k)}, \beta\right) = S\left(x_j^{(2m+1-k)}, \beta\right) = \{k, k+1, k+2, k+3\},$$
where $k = 1, 2, \ldots, m.$

Case 2. $i = 1, 2, \ldots, 2m$, $j = 2, 3, \ldots, n+1$.

It is not hard to see that

$$S\left(x_j^{(k)}, \beta\right) = S\left(x_j^{(2m+1-k)}, \beta\right) = \{k+3j-6, k+3j-5,$$

$k+3j-4, k+3j-3\}$, where $k = 1, 2, \ldots, m$.

Case 3. $i = 1, 2, \ldots, 2m$, $j = n+2, \ldots, 2n-1$.

It is not hard to see that

$$S\left(x_j^{(k)}, \beta\right) = S\left(x_j^{(2m+1-k)}, \beta\right) = \{k-3j+6n+3, k-3j+6n+4,$$

$k-3j+6n+5, k-3j+6n+6\}$, where $k = 1, 2, \ldots, m$.

Case 4. $i = 1, 2, \ldots, 2m$, $j = 2n$.

It is not hard to see that

$$S\left(x_j^{(k)}, \beta\right) = S\left(x_j^{(2m+1-k)}, \beta\right) = \{k+3, k+4, k+5, k+6\},$$

where $k = 1, 2, \ldots, m$.

This shows that $\beta$ is an interval edge $(3n+m)$ – coloring of the graph $G$.
The proof is complete.

From **theorem 2** and **theorem 4** we have the following

**Corollary.** If $G = T(2m, 2n)$, $4 \leq t \leq \max\{3m+n, 3n+m\}$ then $G \in \mathfrak{N}_t$.